\documentclass[aps,twocolumn,showpacs,nofootinbib,superscriptaddress]{revtex4}
\usepackage{graphicx}
\usepackage{amsmath}
\usepackage{amsfonts}
\usepackage{amssymb}
\usepackage{hyperref}

\newcommand{\ket}[1]{|{#1}\rangle}
\newcommand{\bra}[1]{\langle{#1}|}
\newcommand{\traza}{{\rm Tr}}

\begin{document}

\title{Interference in discrete Wigner functions}

\author{Cecilia Cormick}
\affiliation{Departamento de F\'\i sica, FCEyN, UBA,
Ciudad Universitaria Pabell\'on 1, 1428 Buenos Aires, Argentina}

\author{Juan Pablo Paz}
\affiliation{Departamento de F\'\i sica, FCEyN, UBA,
Ciudad Universitaria Pabell\'on 1, 1428 Buenos Aires, Argentina}

\date{\today}

\begin{abstract}

We analyse some features of the class of discrete Wigner functions that was recently introduced by Gibbons \textit{et al.} to represent quantum states of systems with power-of-prime dimensional Hilbert spaces [Phys. Rev. A \textbf{70}, 062101 (2004)]. We consider ``cat" states obtained as coherent superpositions of states with positive Wigner function; for such states we show that the oscillations of the discrete Wigner function typically spread over the entire discrete phase-space (including the regions where the two interfering states are localized). This is a generic property which is in sharp contrast with the usual attributes of Wigner functions that make them useful candidates to display the existence of quantum coherence through oscillations. However, it is possible to find subsets of cat states with a natural phase-space representation, in which the oscillatory regions remain localized. We show that this can be done for interesting families of stabilizer states used in quantum error-correcting codes, and illustrate this by analysing the phase-space representation of the five-qubit error-correcting code. 

\end{abstract}
\date{\today}
\maketitle

\section{Introduction}

Continuous Wigner functions \cite{Wigner32, HilleryOSW84} allow for the study of quantum systems through the representation of their state in phase-space, providing a complete description alternative to the density matrix. The continuous Wigner function $W(q,p)$ is a quasi-probability distribution: it is real valued and by integrating it along phase-space lines one obtains probability densities of certain observables (quadratures). However, it cannot be interpreted as a true probability distribution because it is not positive definite. Negative values are associated with the existence of quantum interference, which is itself identified as a signal of non-classical behaviour \cite{Zurek03,PazZ00}.
The original definition of the Wigner function has been generalized to systems with discrete degrees of freedom \cite{Buot74, HannayB80, CohenS86, Feynman87, Wootters87, GalettiP88, GibbonsHW04}. Among these we find the very relevant case of quantum computers, which are composed by $n$ qubits and therefore have a Hilbert space of dimension $d=2^n$. The discrete analogues of the Wigner function have been recently used to study problems connected with quantum information processing such as the phase-space representation of quantum algorithms \cite{MiquelPS02}, quantum state tomography \cite{Wootters04,PazRS04}, teleportation \cite{KoniorczykBJ01, Paz02}, decoherence in quantum walks \cite{LopezP03}, and the phase-space representation of quantum error-correcting codes \cite{PazRS04b}.

These discrete versions of the Wigner function have a number of peculiar properties which make them rather different from their continuous counterparts. On the one hand, there is not a unique definition and, according to the different proposals, the phase-space can be taken as a $2d\times 2d$ grid \cite{MiquelPS02} or a $d\times d$ one \cite{GibbonsHW04}. Here we shall point out and study some remarkable features of the discrete Wigner functions introduced by Wootters and co-workers \cite{Wootters04,GibbonsHW04} for systems with power-of-prime dimensions, $d=p^n$ (we shall focus our analysis on the case of qubits, $p=2$). These Wigner functions are built on a $d\times d$ grid with coordinates that belong to the finite Galois field $GF(d)$ \cite{LidlN86}. The use of elements of the finite field is the key to the construction of a discrete phase-space whose geometrical properties are similar to those of the continuous phase-space. This geometrical structure is exploited in the definition of $W(q,p)$ and has significant consequences that link the Wigner functions to the stabilizer formalism \cite{Gottesman97} and to the properties of complete sets of mutually unbiased bases \cite{LawrenceBZ02, KlappeneckerR04, BengtssonE04}. As stabilizer states and Pauli operators play crucial roles in the very definition of this class of Wigner functions, these are expected to become a useful tool to analyse some quantum information problems, such as those related to stabilizer codes, Pauli noise channels, etc \cite{PazRS04b, cormick-2006-73}.

Some of our results indicate a potential problem for the class of Wigner functions defined by Wootters and co-workers. For the sake of clarity we describe this problem here in simple terms: these Wigner functions might not always be a powerful tool to pictorially display properties of quantum states. This is in contrast to the continuous Wigner function, which is often helpful to put quantum interference on display and to study, for example, its disappearance due to decoherence \cite{LopezP03}. It is known that certain states have positive Wigner functions and are often referred to (in a loose way) as ``classical states". For the discrete Wigner functions such classical states have been studied elsewhere 
\cite{cormick-2006-73, gross-2006-} and belong to the class of stabilizer states.
As expected, Wigner functions of states which are coherent superpositions of ``classical" states are not positive, but display oscillations. In the usual continuous version of the Wigner function the oscillations are localized in phase-space and concentrate in the region between the two classical states. 
The discrete Wigner functions we study behave in a drastically different way: in general the interference spreads over all phase-space, affecting even the regions where the original ``classical'' states are localized. We investigate this property in detail and find that it is a generic feature. For example, we show that as the number of qubits increases the interference tends to spread uniformly over all phase-space. Simultaneously, the ratio between the typical value of the interference terms and that of the ``classical" terms decays exponentially. As a consequence, the presence of interference becomes harder to identify when using this class of Wigner functions.

However, the discrete Wigner function proposed by Wootters and co-workers can be defined in a number of ways. Exploiting this freedom we show that given two orthogonal stabilizer (classical) states it is possible to define a Wigner function such that all coherent superpositions of those states have a natural phase-space representation (i.e., one in which the quantum interference is localized). This class of states contains some cases that are very interesting from the quantum information point of view. For example, stabilizer quantum error-correcting codes \cite{Shor95,Steane96,Gottesman00} define encoded states as arbitrary superpositions of certain stabilizer states (the logical states). We analyse in detail the five-qubit code (that corrects all one-qubit errors and encodes one qubit of information \cite{LaflammeMPZ96}) and show how to define a suitable Wigner function to represent it. The phase-space representation of this class of error-correcting codes is interesting because Pauli errors correspond to phase-space translations and, therefore, have also a natural interpretation. 

The paper is organized as follows: in Section \ref{sec:wignerdef} we shortly review the definition of the discrete Wigner functions. Section \ref{sec:interfwigner} contains a derivation of the expressions for the Wigner functions of arbitrary superpositions of two computational states, and a brief analysis of some of their features. In Section \ref{sec:overlap} we discuss the conditions the Wigner function should satisfy to avoid the overlap between the interference and the interfering computational states. In Section \ref{sec:average} we study the average behaviour of the interference fringes. Finally, in Section \ref{sec:5qubitcode} we discuss the phase-space representation of the five-qubit error-correcting code. In Section \ref{sec:conclusion} we summarize our results.

\section{Discrete Wigner functions} 
\label{sec:wignerdef}

We present here a short description of the Wigner functions proposed in \cite{GibbonsHW04}, restricting our attention to systems with a space of states with dimension $d=2^n$ and skipping many details and technicalities that can be found in \cite{GibbonsHW04, PazRS04b}.
The first step in the construction of the discrete Wigner function is the choice of the phase-space. This is taken as a $d \times d$ grid with ``position'' and  ``momentum'' coordinates $(q, p)$ that are elements of the finite field $GF(d)$, which contains a zero element and the $d-1$ different powers $\omega^j$, $j=0, \ldots d-2$, of a generating element $\omega$. Lines in phase-space are defined as sets of points $\alpha = (q, p)$ that solve linear equations of the form $aq+bp=c$, where all elements and operations are in $GF(d)$ (for instance, all sums are modulo 2). The field structure is responsible for the validity of the following geometrical properties of the phase-space: two lines are either parallel (i.e., they do not intersect) or they share exactly one point; given a line there are $d-1$ lines which are parallel to it (the $d$ parallel lines form a ``striation''); every line has exactly $d$ points and there are exactly $d+1$ different striations.

\begin{figure}[!hbt]
\begin{center}
\includegraphics[width=0.16\textwidth]{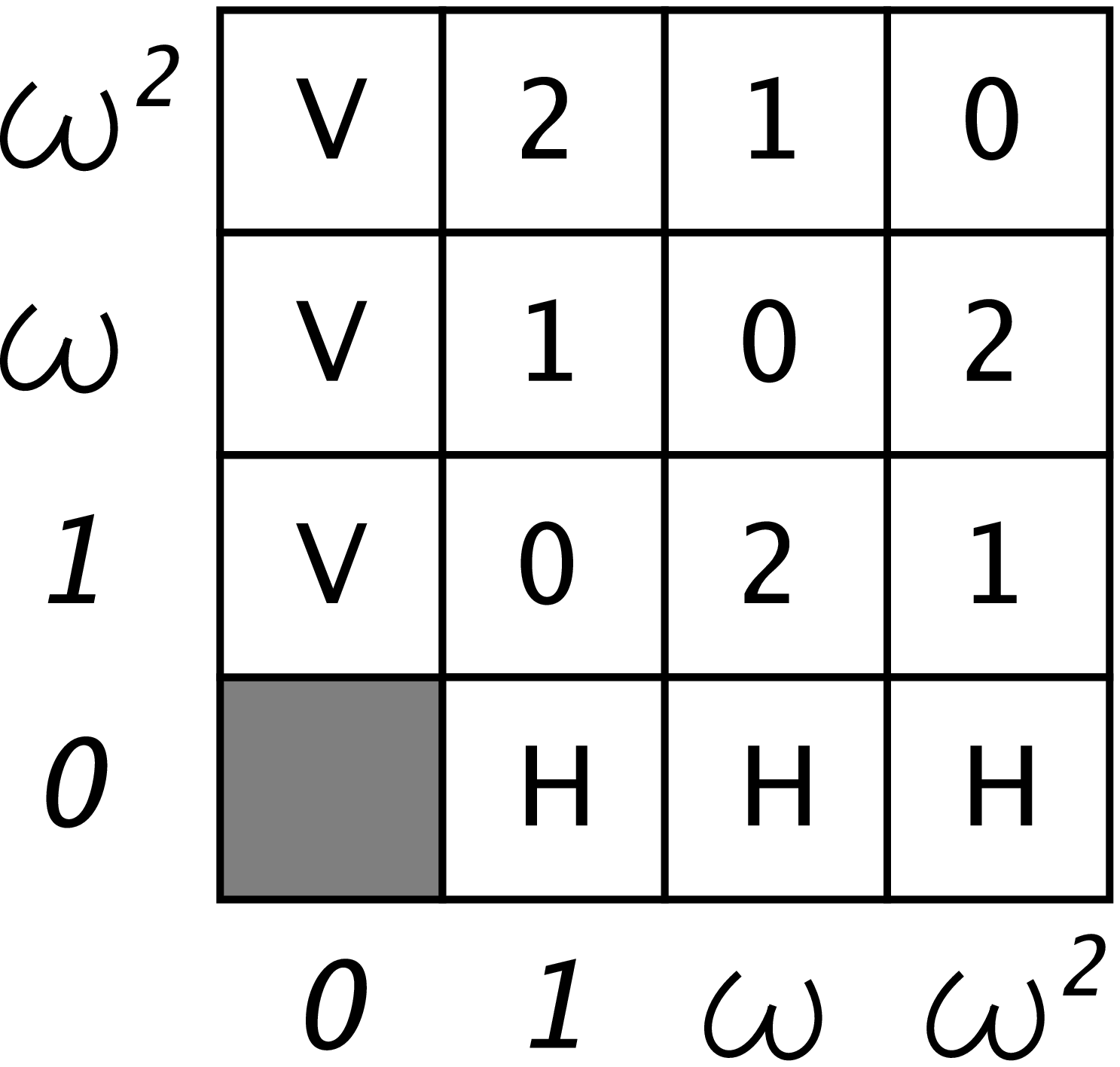}
\caption{The phase-space for a system of two qubits. The five different rays are indicated: one is horizontal ($p=0$, labelled H), one vertical ($q=0$, labelled V), and the other three oblique ($\omega^k q+p=0$, labelled by the parameter $k$). There is one ray in each striation, and all rays cross the origin.}
\label{fig:2qubitrays}
\end{center}
\end{figure}

The following notational conventions will be useful: the symbol $\lambda$ will refer to a line in phase-space; $\lambda_\alpha$ will be a line that contains the point $\alpha$, and $\lambda^{(\kappa)}$ a line that belongs to the striation $\kappa$. Accordingly, $\lambda^{(\kappa)}_\alpha$ will denote the only line in the striation $\kappa$ that contains the point $\alpha$, and $\lambda_{\alpha, \beta}$ the only line that joins the points $\alpha$ and $\beta$. The lines of the form $\lambda_{0, \beta}$ are lines which contain the origin, and will be called ``rays"; there is one ray in each of the striations. In Figure \ref{fig:2qubitrays} we show the phase-space for a two-qubit system, indicating the five different rays.
\smallskip

The key to the construction of the Wigner function is to associate each line $\lambda$ to a rank-one projection operator $P(\lambda)$. The Wigner function $W(\alpha)$ is then required to satisfy certain conditions: first, the sum of its values along the line $\lambda$ must be equal to the probability associated to the projector $P(\lambda)$: 
\begin{equation}
\sum_{\alpha \in \lambda} W(\alpha) = \traza \Big(\rho P(\lambda)\Big)
\label{ECwignermed}
\end{equation}
where $\rho$ is the state of the system. Also, in analogy with the continuous case, the inner product of two states must be obtained from the Wigner function as: 
\begin{equation}
\traza (\rho_1 \rho_2) = d \sum_\alpha W_1(\alpha) W_2(\alpha)
\label{ECwignerprod}
\end{equation} 
Because of the phase-space geometry, these conditions imply that the $d$ lines in a given striation should be associated to $d$ projectors corresponding to states in an orthonormal basis (which means the Wigner function is normalized to one). Moreover, we impose that the Wigner function of the ``line state" $P(\lambda)$ is equal to zero everywhere except from the line $\lambda$ (where it is constant and equal to $1/d$). Then, as lines belonging to different striations intersect at a single point, the bases associated to different striations must be mutually unbiased \cite{LawrenceBZ02}. 

The conditions (\ref{ECwignermed}, \ref{ECwignerprod}) determine that the Wigner function at any phase-space point is given by:
\begin{equation}
W(\alpha) = \frac{1}{d} \left\{ \sum_\kappa \traza \Big( \rho P(\lambda^{(\kappa)}_\alpha) \Big) - 1 \right\}. 
\label{ECwignerdef}
\end{equation}
Thus, $W(\alpha)$ depends on the probabilities for all the states related to the lines (one in each striation) which contain the point $\alpha$. As in the continuous case, it is convenient to express the Wigner function as the expectation value of the so-called ``phase-space point-operators" $A(\alpha)$:
\begin{eqnarray}
W(\alpha)&=&{\rm Tr}\left(\rho A(\alpha)\right)\\
A(\alpha)&=&\frac{1}{d}\left\{ \sum_\kappa P(\lambda^{(\kappa)}_\alpha) - \mbox{$1
\hspace{-1.0mm}  {\bf l}$}  \right\} 
\label{eqpointop}
\end{eqnarray}
These point-operators are hermitian and form an orthonormal basis for the space of operators (using the Hilbert-Schmidt inner product).

To complete the construction above and define the Wigner function it is necessary to provide a mapping between lines in phase-space and states in Hilbert space. We impose over this mapping some additional geometric constraints. Let us define a set of $d\times d$ unitary operators $T(q,p)$ acting on the Hilbert space that faithfully represent discrete phase-space translations. By this we mean that: i) one such operator is associated to each phase-space point, and ii) the product of two such operators is, up to a phase, the operator associated to the sum of the corresponding phase-space points. It is worth noticing that this implies that the set of $d\times d$ unitaries must be closed under the product (up to a phase) and  form a group. For the association between lines and states to respect covariance under translations, the quantum state associated to a translated line should be identical to the state obtained by acting with the operator $T(q,p)$ on the original line state. Also, all the lines in a certain striation remain invariant under translations associated to the points located in the ray belonging to such striation (i.e, the line satisfying the equation $aq+bp=c$ is mapped onto itself under the transformation $q'=q+q_0$, $p'=p+p_0$ when $aq_0+bp_0=0$). This implies that the basis associated to a certain striation should be formed by the common eigenstates of the translation operators associated to the points in the ray. For this to be possible, of course, such operators must commute. So the phase-space construction must be such that the operators in the rays define $d+1$ mutually unbiased orthogonal bases. 

It is interesting to notice that the idea of representing quantum states in the phase-space chosen naturally requires the splitting of a group of $d\times d$ operators into $d+1$ Abelian subsets defining $d+1$ mutually unbiased bases in Hilbert space. The elements of the Pauli group are natural candidates to be used as translation operators. These generalized Pauli operators are defined in terms of two binary n-tuples $\vec q$ and $\vec p$ as:
\begin{equation}
T(\vec q, \vec p) = \prod_{j = 1}^n X_{(j)}^{q_j} Z_{(j)}^{p_j} e^{i\frac{\pi}{2} q_j p_j} \equiv X^{\vec q} Z^{\vec p} e^{i\frac{\pi}{2} \vec q \cdot \vec p}.
\label{ec:translations}
\end{equation}
Here the subscripts indicate on which qubit the operators act, the binary $n$-tuples $\vec q$, $\vec p$ determine which Pauli matrix acts on each of the qubits, and the phase is chosen to make the operators hermitian. 

\begin{figure}[!hbt]
\begin{center}
\includegraphics[width=0.38\textwidth]{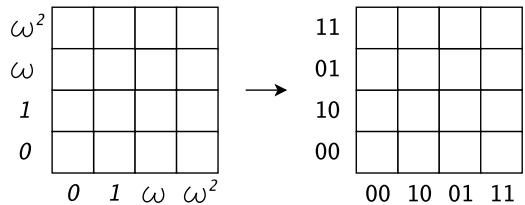}
\caption{The assignment of $n$--tuples to the phase-space coordinates for a system of two qubits; this allows to associate to each point a Pauli operator $T(\vec q, \vec p)$ that represents a translation in phase-space.}
\label{fig:2qubitcoord}
\end{center}
\end{figure}

To associate a Pauli operator to each phase-space point we must map the $n$--tuples ($\vec q$, $\vec p$) to the field elements $(q,p)$. This must be done in a way that ensures the commutativity of the operators $T(\vec q, \vec p)$ that belong to the same ray. The simplest way to find $n$--tuples satisfying this constraint is using matrix representations of the field $GF(d)$, i.e. a set of $d-1$ non-zero matrices which are obtained as the first $d-1$ powers of a generating binary matrix $M$ (plus the zero matrix). Using such set, one can assign the position and momentum $n$--tuples by the following prescription: first, we associate the zero $n$--tuple to the zero element of the field. The $n$--tuples associated to the non-zero field elements can be taken as $q_j=\omega^{j-1}\rightarrow \vec q_0 M^{j-1}$, $p_j=\omega^{j-1}\rightarrow \vec p_0 \tilde M^{j-1}$, where $j=1,\ldots,d-1$, $\tilde M$ is the transpose of $M$ and $\vec q_0$, $\vec p_0$ are arbitrary $n$--tuples \cite{PazRS04b}. Given this assignment one can show that the Pauli operators located along the same ray commute (the Paulis associated to the ray $\omega^k q+p=0$ are generated by the operators $T(\vec q_0 M^j, \vec p_0 \tilde M^{j+k})$ for $j=1,\ldots n$). This method is not only a way to build a consistent phase-space structure but also a constructive way to split the Pauli group into $d+1$ Abelian subgroups. In Figure \ref{fig:2qubitcoord} we illustrate the case of two qubits, where $M$ is a symmetric matrix and therefore the assignment of $n$--tuples can be chosen to be the same for the position and momentum coordinates.
\smallskip

Once the coordinates are mapped to $n$--tuples according to this procedure, the association between lines and states must be such that the Wigner function exhibits covariance under translations:
\begin{equation}
\left\{
\begin{array}{ccc}
\rho & \to & T(\vec q_0, \vec p_0) ~\rho~ T^\dagger(\vec q_0, \vec p_0)\\
\\
W(\vec q, \vec p) & \to & W(\vec q + \vec q_0, \vec p + \vec p_0)  
\end{array}
\right.
\label{ECcovarianciatras}
\end{equation}

The method described has an important ambiguity (which can be interpreted also as an important degree of freedom): it assigns a basis of Hilbert space entirely formed by stabilizer states to every  phase-space striation, but it does not determine which state in the basis we should associate to each line in the striation. As discused in \cite{GibbonsHW04}, the freedom in this choice leads to a number of possible ``quantum nets'', each of which is a mapping $\lambda \to P(\lambda)$ between lines in phase-space and states in the corresponding bases. The definition of the Wigner function will depend on the quantum net chosen, so the recipe we have outlined results in a class of different possible Wigner functions. At the moment there is no criterion to choose a unique definition, though there have been attempts to reduce the freedom in the choice of the quantum net \cite{PazRS04b, PittengerR05}.

Finally, a few notational remarks. It will be convenient to express the point-operators in terms of the Pauli operators in the way \cite{PazRS04b}:
\begin{equation}
A(\alpha) = \frac{1}{d^2} \sum_\beta (-1)^{\alpha \wedge \beta} f_\beta ~T(\beta)
\label{ECfbeta-A}
\end{equation}
where $\alpha \wedge \beta = \vec q_\alpha \cdot \vec p_\beta - \vec q_\beta \cdot \vec p_\alpha$ is the symplectic product between $\alpha = (\vec q_\alpha, \vec p_\alpha)$ and $\beta = (\vec q_\beta, \vec p_\beta)$, and the $f_\beta$ are constants given by:
\begin{equation}
\left\{ 
\begin{array}{ll}
f_0=1, \\
f_\beta = \traza \Big( P(\lambda_{0, \beta}) T(\beta) \Big), \quad & \beta \neq 0
\end{array} \right.
\label{eq:f_beta}
\end{equation} 
Because of the covariance condition, the projectors $P(\lambda_{0, \beta})$ are eigenstates of the Pauli operators $T(\beta)$, and so the constants $f_\beta$ are the corresponding eigenvalues. Each of them is then equal to $\pm 1$, and the choice of the set $\{f_\beta\}$ completely fixes the quantum net. But these constants are not all independent; for each projector $P(\lambda_{0, \beta})$ we can choose only $n$ eigenvalues, corresponding to the $n$ Paulis that generate the set of $d$ Paulis on the ray of the striation.

\section{Interference in the discrete Wigner function}
\label{sec:interfwigner}

Given two states $\ket{\psi_1}$ and $\ket{\psi_2}$ with Wigner functions $W_1(\alpha)$ and $W_2(\alpha)$, the Wigner function for the superposition state $\ket{\psi} = a \ket{\psi_1} + b \ket{\psi_2}$ is:
\begin{eqnarray}
W(\alpha) & = & |a|^2 W_1(\alpha) + |b|^2 W_2(\alpha) + \nonumber \\
& & + 2 \mathcal{R}e \{a b^* \bra{\psi_2} A(\alpha) \ket{\psi_1} \}.
\end{eqnarray}
Thus, the Wigner function of the coherent superposition is a weighted sum of the Wigner functions of the states $\ket{\psi_1}$ and $\ket{\psi_2}$, plus an interference term. If we consider states $|\psi_{1,2}\rangle$ such that their Wigner function is non-negative (line states, for the discrete case discussed above \cite{cormick-2006-73}), the interference term is responsible for the negativity of the total Wigner function. In the continuous case, it is well known that the only pure states with non-negative Wigner functions are the coherent states, whose representations in phase-space are Gaussian. Superpositions of such states have Wigner functions with oscillations which are localized in the phase-space region between the two Gaussian peaks. In what follows we shall see that for the class of discrete Wigner functions with coordinates in $GF(d)$ sketched in the previous section the representation of interference can be drastically different.

We consider a system of $n$ qubits and focus on the simplest case: a superposition of two states in the computational basis. This corresponds to the superposition of a pair of vertical lines, and can be easily generalized to superpositions of states associated to any pair of parallel lines. 
States in the computational basis are written as: $\ket{\vec k} = \ket{k_1 \ldots k_n}$, with $\vec k$ a binary $n$-tuple. We shall take $\ket{\psi_1} = \ket{\vec 0}$, $\ket{\psi_2} = \ket{\vec m}$ ($\vec m \neq \vec 0$); any other combination of computational states can be obtained by translation of states of the form considered. The Wigner function $W_1$ ($W_2$) takes the value $1/d$ on the points that belong to the vertical line $\vec q = \vec 0$ ($\vec q = \vec m$), and vanishes everywhere else. The interference term can be calculated from:
\begin{eqnarray}
\bra{\vec m} A(\vec q, \vec p) \ket{\vec 0} & = & \frac{1}{d^2} \sum_{\vec q~', \vec p~'} (-1)^{\vec q \cdot \vec p~' - ~\vec q~' \cdot \vec p} ~f_{(\vec q~', \vec p~')} \nonumber \\
& & \bra{\vec m} T(\vec q~', \vec p~') \ket{\vec 0} = \\
\smallskip \nonumber \\
 & = & \frac{(-1)^{\vec m \cdot \vec p}}{d^2} \sum_{\vec p~'} f_{(\vec m, \vec p~')} (-1)^{\vec q \cdot \vec p~'} e^{i \frac{\pi}{2} \vec m \cdot \vec p~'} \nonumber 
\label{ECinterfcomp}
\end{eqnarray}
According to this expression, the interference term depends on the position and momentum coordinates in a separable form. The only dependence on the momentum $\vec p$ is given by an oscillation with ``frequency'' determined by the displacement $\vec m$ between the lines in the superposition. The dependence on the position coordinate is more involved and is given by:
\begin{equation}
F(\vec q) = \frac{1}{d^2} \sum_{\vec p~'} f_{(\vec m, \vec p~')} (-1)^{\vec q \cdot \vec p~'} e^{i \frac{\pi}{2} \vec m \cdot \vec p~'}
\label{ec:laF(q)}
\end{equation} 
which depends also on the relative displacement $\vec m$ and on the quantum net chosen.

In contrast with the behaviour of interference in the continuous Wigner function, here the interference term may be non-zero at any point in phase-space. For any fixed pair of states $\ket{\vec 0}$ and $\ket{\vec m}$, it is possible to find a quantum net for which the Wigner function mimics (as much as possible) the continuous case, being non-zero only on a pair of vertical lines. This can be done by imposing the condition that the sum (\ref{ec:laF(q)}) be totally constructive at some arbitrary value $\vec q_I$; then, as $F(\vec q + \vec m) = F(\vec q)^*$, it will be constructive at $\vec q_I + \vec m$ too, and it will vanish at any other value of $\vec q$.

\begin{figure}[!hbt]
\begin{center}
\includegraphics[width=0.4\textwidth]{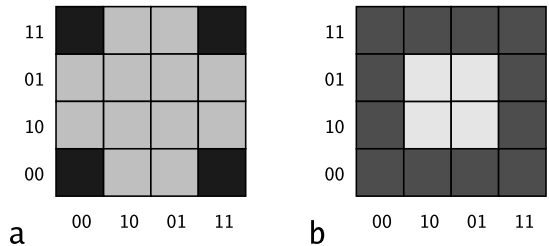}
\caption{The possible Wigner functions for the Bell state $\ket{\phi_+} = (\ket{00} + \ket{11}) / \sqrt{2}$, which is a superposition of two computational states. a) For some choices of the quantum net, the Wigner function is nonzero only at the corners of the phase-space; this happens because the interference term overlaps the interfering lines. b) For the remaining quantum nets, the Wigner function is positive at the boundaries of phase-space and negative at the central points; this corresponds to having the constant original lines and the interference lying between them.}
\label{FIGwignerphi+}
\end{center}
\end{figure}

As an example, let us look at the Bell state $\ket{\phi_+} =$ $= (\ket{00}+\ket{11}) / \sqrt{2}$. This state is a superposition of the computational states $\ket{00}$ and $\ket{11}$, and is an eigenstate of the generalized Pauli operators $X_{(1)}X_{(2)}$ and $Z_{(1)}Z_{(2)}$. This implies that its Wigner function must be invariant under the translations in phase-space associated to these operators. In \cite{PazRS04b} it is argued that, taking these symmetries into account, only two possible representations can be obtained for this state; these are shown in Figure \ref{FIGwignerphi+}. In one of them (Fig \ref{FIGwignerphi+}.a) the Wigner function is nonzero only at the corner points; this corresponds to having the interference right over the interfering vertical lines. In the other representation (Fig \ref{FIGwignerphi+}.b) the Wigner function is negative at the central points of phase-space and positive everywhere else; this is what we get when the interference lies between the original lines.

\smallskip

This case shows two qualitatively different ways in which interference can behave: it can lie between the lines in the superposition, or overlap them; the two possible representations look very different from one another.

\section{The overlap between the interference and the original lines cannot be avoided}
\label{sec:overlap}

In this section we will analyse the conditions the quantum net should satisfy so that the interference always lies outside the interfering vertical lines. This means that $F(\vec q)$ (\ref{ec:laF(q)}) should be zero at $\vec q = \vec 0$ (and then it will vanish at $\vec q = \vec m$ too). This requirement is given by:
\begin{equation}
F(\vec 0) = \frac{1}{d^2} \sum_{\vec p~'} f_{(\vec m, \vec p~')} e^{i \frac{\pi}{2} \vec m \cdot \vec p~'} = 0
\label{ECinterffuera}
\end{equation} 

The previous equation can be separated in its real and imaginary parts; for each value of $\vec m$ we then have two linear equations on the set $f_{(\vec m, \vec p~')}$. If we want the interference not to overlap the original lines for arbitrary pairs of vertical lines, we have a total of $2(d-1)$ equations, for the $d-1$ non-zero values of $\vec m$. But the constants $f_{(\vec q, \vec p)}$ are not all independent: we can choose arbitrarily the values for $f_{(\vec e_i, \vec p)}$, where $(\vec e_i)_j = \delta_{ij}$ for $i,j = 1, \ldots, n$; the remaining values are then fixed, and can be obtained as products of the independent ones. So the conditions on the quantum net are non-linear and become more involved as the number of qubits is increased. In the following we shall focus on the cases of two and three qubits, and see that no quantum net can satisfy the whole set of equations for all $\vec m$. 

For a system of two qubits, there are three non-zero $2$-tuples $\vec m$. For each of them, we can write down the conditions (\ref{ECinterffuera}) as:
\begin{equation}
\left\{ \begin{array}{ccc}
f_{(10, 00)} + f_{(10, 01)} + i (f_{(10, 10)} + f_{(10, 11)}) & = & 0 \\
f_{(01, 00)} + f_{(01, 10)} + i (f_{(01, 01)} + f_{(01, 11)}) & = & 0 \\
f_{(11, 00)} - f_{(11, 11)} + i (f_{(11, 10)} + f_{(11, 01)}) & = & 0 
\end{array} \right.
\label{ECinterffuera2}
\end{equation}
The constants $f_{(\vec q, 00)}$ can be chosen to be 1 without any loss of generality \cite{GibbonsHW04}.
As mentioned before, the other constants are not independent: the non-zero points on each ray are related to a set of three commuting Paulis generated by two of them, so that the third Pauli in each ray is (up to a sign) a product of the other two. Taking into account the phases of the Paulis $T(\vec q, \vec p)$ (\ref{ec:translations}) we can see that:
\begin{equation}
\left\{ \begin{array}{ccr}
T(11, 10) & = & - T(10, 01) ~T(01, 11)\\
T(11, 01) & = & - T(10, 11) ~T(01, 10)\\
T(11, 11) & = & T(10, 10) ~T(01, 01)
\end{array} \right.
\end{equation}
As the $f$'s are the eigenvalues of the translation operators (\ref{eq:f_beta}), they must satisfy the same relations:
\begin{equation}
\left\{ \begin{array}{ccr}
f_{(11, 10)} & = & - f_{(10, 01)} f_{(01, 11)}\\
f_{(11, 01)} & = & - f_{(10, 11)} f_{(01, 10)}\\
f_{(11, 11)} & = & f_{(10, 10)} f_{(01, 01)}
\end{array} \right.
\end{equation}

Plugging this in the conditions (\ref{ECinterffuera2}) we find that:
\begin{equation}
\left\{ \begin{array}{ccc}
f_{(10, 10)} + f_{(01, 01)} & = & 0 \\
f_{(10, 10)} f_{(01, 01)} & = & 1
\end{array} \right.
\end{equation}
which cannot be solved by any choice of $f_{(10, 10)}$ and $f_{(01, 01)}$. Thus, for a system of two qubits there is no quantum net that avoids the overlap for arbitrary pairs of computational states.

As the number of qubits in the system is increased, the problem becomes much more difficult, not only because of the exponential number of equations, but also because the maximum degree of the equations is given by the number of qubits. For the case of three qubits, we wrote the conditions (\ref{ECinterffuera}) and the relations between the constants $f$, obtaining a system of 14 non-linear equations for the 21 independent constants. We examined all the possible sets of values and none of them solved the system. In this way, we found that no quantum net can avoid the overlap in the three-qubit case.

From the results corresponding to systems of two and three qubits, it seems a natural conjecture that for higher numbers of qubits it will not be possible either to choose a quantum net such that the interference lies outside the interfering lines for arbitrary superpositions of states in the computational basis.

\section{Average behaviour of interference in the Wigner function}
\label{sec:average}

As no quantum net seems to be particularly convenient to represent arbitrary superpositions of computational states, we studied the ``average" representation of interference by picking the quantum nets at random. Our goal was to analyse its behaviour for large numbers of qubits.

As was pointed out in section \ref{sec:interfwigner}, for each value of $\vec m$ we can always find a quantum net such that the interference concentrates in two vertical lines (with relative displacement $\vec m$). In this case, over those two lines the real and imaginary parts of $F(\vec q)$ satisfy $|\mathcal{R}e \left( F(\vec q) \right)| =$ $= |\mathcal{I}m \left( F(\vec q) \right)| = 1/2d$; this means that the values of the interference term are of the same order as those of the Wigner function on the interfering lines. However, if we choose a different value for $\vec m$ and keep the same quantum net, the interference will not retain this nice behaviour, and it will spread and fade out. 

We studied the ``average'' properties of interference in the following way: we calculated the distributions $F(\vec q)$ for random quantum nets, and we worked separately with the absolute values of the real and imaginary parts, defining the new  functions $R(\vec q) = |\mathcal{R}e \left( F(\vec q) \right)|$, $I(\vec q) = |\mathcal{I}m \left( F(\vec q) \right)|$. We found the maxima for these two functions, so as to compare the order of magnitude of the interference with the value of the Wigner function on the original lines (which is of order $1/d$). Besides, we calculated the entropy associated to the distributions $R(\vec q)$, $I(\vec q)$. In order to do so, they were normalized to one, so that their values could be interpreted as probabilities $\mathcal{P}(\vec q)$. To facilitate the comparison between systems with different dimensions, we defined a normalized entropy given by:
\begin{equation}
S = - \sum _{\vec q} \mathcal{P}(\vec q) ~log_d \left( \mathcal{P}(\vec q) \right)
\end{equation}
Then, the entropy for a distribution concentrated in only one value of $\vec q$ is equal to zero, while the maximal entropy corresponds to the uniform probability distribution, and takes a value of one for any number of qubits. 

\begin{figure}[!hbt]
\begin{center}
\includegraphics[width=0.4\textwidth]{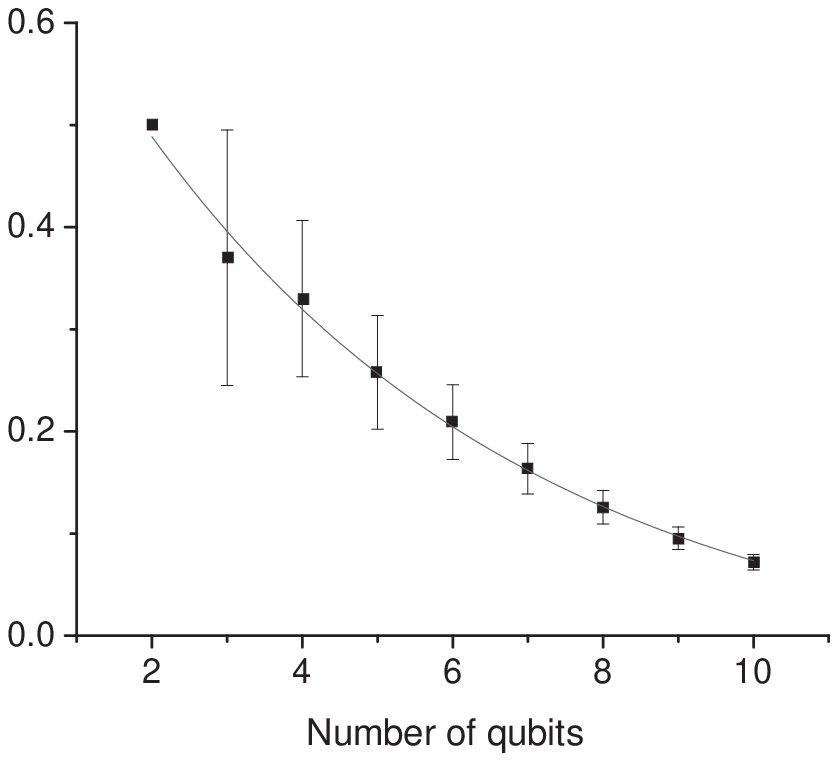}
\includegraphics[width=0.39\textwidth]{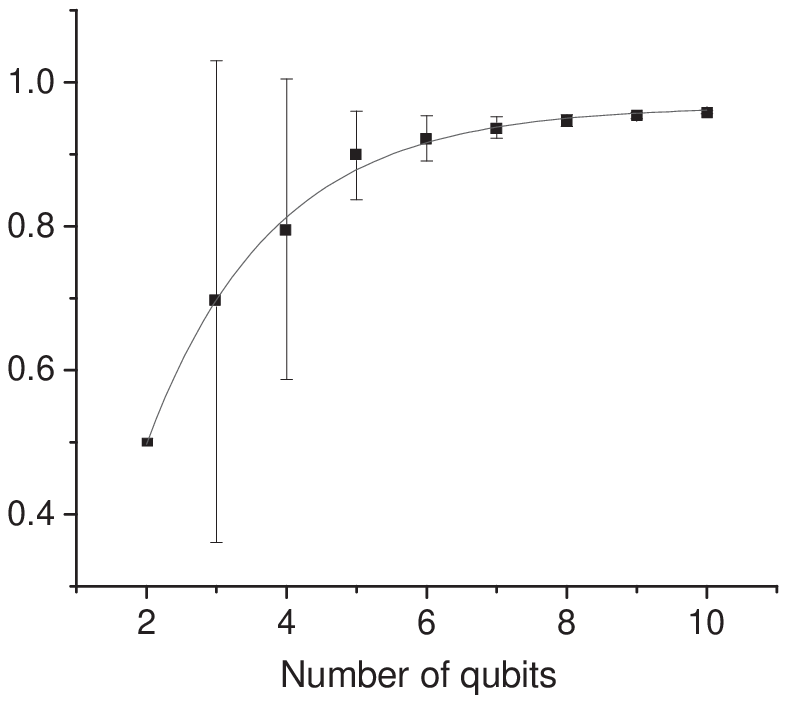}
\caption{``Average'' properties of interference in phase-space. On top: decay of the ratio between the maximal interference values and those of the Wigner function on the original lines. Below: normalized entropy of the interference distribution. The results are averaged over 50 random quantum nets and the $d-1$ non-zero relative displacements $\vec m$, and the error bars are given by the mean deviations. The continuous line shows the exponential fit.}
\label{FIGinterf}
\end{center}
\end{figure}

Maximal values and entropies for the distributions $R(\vec q)$ and $I(\vec q)$ were calculated for 50 random quantum nets, and for each possible displacement $\vec m$, in systems of 2 to 10 qubits. For each number of qubits, the $50 (d-1)$ values were averaged. The results for $R(\vec q)$ and $I(\vec q)$ were the same, taking into account their mean deviations. We found that the ratio between the average of the maximal interference values and those of the Wigner function on the original lines decays exponentially in the number of qubits. For the entropy we found a saturation trend as the system is enlarged; this means that for large numbers of qubits the interference is spread over the whole phase-space taking almost uniform absolute values. As a consequence of these two results (shown in Figure \ref{FIGinterf}), interference becomes hard to detect using these Wigner functions for systems composed by more than a few qubits. 

\smallskip

\section{A Wigner function analysis of the five-qubit error-correcting code}
\label{sec:5qubitcode}

There are important cases in which we are not interested in arbitrary superpositions of all possible pairs of states in a given basis. For example, in a stabilizer quantum error-correcting code the encoded states are superpositions of a subset of stabilizer states. If a single qubit is encoded, a general state is a superposition of a fixed pair of stabilizer states. In such case, we can always find a suitable definition of the Wigner function, in which all the encoded states have simple representations. Below, we will show how to implement and use this construction to represent the ``perfect" five-qubit code, that encodes one qubit of information and corrects all one-qubit errors \cite{LaflammeMPZ96}. 

\subsection{The perfect five-qubit code}

This code protects the state of one qubit by encoding it in a five-qubit system. The space of ``legal" encoded states is formed by all the linear combinations of the logical states $\ket{0}_L$ and $\ket{1}_L$ (which are the encoded one-qubit computational states). The states $\ket{0}_L$ and $\ket{1}_L$ are stabilizer states; more precisely, the state $\ket{0}_L$ can be chosen to be the graph-state \cite{SchlingemannW01, Schlingemann02} associated to a five-qubit ring.
The stabilizer of this graph is generated by the Pauli operators $S_j$ that apply $X$ on the $j$-th vertex and $Z$ on all its neighbours, 
\begin{equation}
S_j = Z_{(j-1)} X_{(j)} Z_{(j+1)}
\end{equation}
(where indices are cyclic from 1 to 5). The subspace of encoded states is formed by the states that are stabilized by all the operators of the form $S_j S_k$. The graph-state, which is stabilized by the operators $S_j S_k$ and by the operator $\widetilde S = S_1 \ldots S_5 = -X_{(1)} \ldots X_{(5)}$, is chosen as the logical state $\ket{0}_L$. The state $\ket{1}_L$ is then taken to be the state which is stabilized by the operators $S_j S_k$, and by the operator $-\widetilde S$. This means that $\widetilde S$ plays the role of the encoded Pauli operator $Z$, that distinguishes between the two computational states. The encoded computational states are connected by the encoded $X$ operator:
\begin{equation}
\ket{1}_L = Z_{(1)} \ldots Z_{(5)} \ket{0}_L
\label{ec:compstates}
\end{equation}

\subsection{A convenient definition for the Wigner function}

To represent this stabilizer code we can use an alternative definition of the Wigner function, that can be built following the same procedure as before but redefining the translations in Hilbert space. This will lead us to a Wigner function with the same geometrical structure, but with a particularly simple representation for the states we want to study. An essential step in the definition of the Wigner function is the identification of the translations with the generalized Pauli operators (\ref{ec:translations}). In the previous sections we chose to represent the vertical translations with tensor products of $Z$ Pauli matrices and the horizontal translations with tensor products of $X$ matrices. For this reason, the vertical lines were associated to the computational basis while the horizontal lines corresponded to the eigenstates of all $X$ operators. However, we can choose our vertical and horizontal translations in a different way, still using Pauli operators. In order to maintain the same structure for the discrete Wigner function, we only need to find sets of generalized Pauli operators $X'_j$ and $Z'_j$ ($j = 1, \ldots, n$) with the same commutation-anticommutation relations as the $X_{(j)}$, $Z_{(j)}$ (note that the $X'_j$ and $Z'_j$ may act on more than one qubit). Then we can redefine the translations as:
\begin{equation}
T'(\vec q, \vec p) = \prod_{j = 1}^n {X'_j}^{q_j} {Z'_j}^{p_j} e^{i\frac{\pi}{2} q_j p_j} = {X'}^{\vec q} {Z'}^{\vec p} e^{i\frac{\pi}{2} \vec q \cdot \vec p}
\label{ECtrasalter}
\end{equation}

The Wigner function is now defined as the expectation value of the new point operators, given by:
\begin{equation}
A'(\alpha) = \frac{1}{d^2} \sum_\beta (-1)^{\alpha \wedge \beta} f'_\beta ~T'(\beta)
\end{equation}
(where the $f'$s are eigenvalues of the redefined translations).
The construction proceeds just the same way as before; each striation in phase-space will be related to a stabilizer basis, and the different bases will still be mutually unbiased, but the basis that corresponds to each of the striations might change. To represent the states in the perfect five-qubit code it is convenient to choose the generators of vertical and horizontal translations as:
\begin{eqnarray}
&& X'_j = S_j = Z_{(j-1)} X_{(j)} Z_{(j+1)} \\
&& Z'_j = Z_{(j)}
\end{eqnarray}
Then, the vertical translations are still products of $Z$ operators (and so the vertical lines are still the computational states) but the horizontal translations are associated to the stabilizers of the graph-state. In this case, the horizontal ray can be chosen to be the graph-state (the encoded $|0\rangle$ state), and by applying $Z$ operators to it we can obtain the states associated to the remaining horizontal lines.

\subsection{The representation of the encoded states}

As we mentioned above, the Wigner function we have just defined is such that the computational state $\ket{0}_L$ has a particularly simple representation: it corresponds to the horizontal ray, $\vec p = \vec 0$. So its Wigner function takes the value $1/d$ on that line, and vanishes everywhere else. The other computational state, $\ket{1}_L$, can be obtained from equation (\ref{ec:compstates}), and is therefore associated to the horizontal line $\vec p = (1,1,1,1,1) = \vec m$. All the encoded states are superpositions of these two ``horizontal'' states.

Following steps similar to those in Section \ref{sec:interfwigner}, we can write down the Wigner function for the state $\ket{\psi}_L=$ $=a\ket{0}_L + b \ket{1}_L$ as:
\begin{equation}
W_\psi'(\vec q, \vec p) = |a|^2 W_0'(\vec q, \vec p) + |b|^2 W_1'(\vec q, \vec p) + W_{I}'(\vec q, \vec p)
\label{ECwignercode}
\end{equation}
with:
\begin{eqnarray}
W_0'(\vec q, \vec p) & = & (1/d)~ \delta(\vec p) \\
W_1'(\vec q, \vec p) & = & (1/d)~ \delta(\vec p - \vec m) \\
W_{I}'(\vec q, \vec p) & = & (2/d^2)~ (-1)^{\vec q \cdot \vec m} \nonumber \\
			& &\mathcal{R}e \Big\{a b^* \sum_{\vec x} (-1)^{\vec x \cdot \vec p} f'_{(\vec x, \vec m)} e^{-i\frac{\pi}{2} \vec x \cdot \vec m} \Big\} \quad
\end{eqnarray}
where the last sum is over all binary $n$-tuples $\vec x$. So, the Wigner function is a superposition of the horizontal lines associated to the states $\ket{0}_L$ and $\ket{1}_L$, plus an interference term that oscillates along the horizontal direction and depends on the quantum net chosen. 

\begin{figure}[!hbt]
\begin{center}
\includegraphics[width=0.45\textwidth]{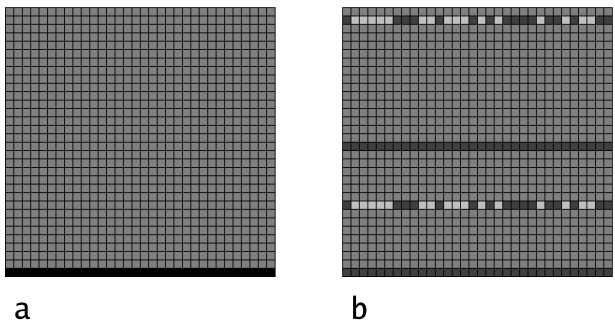}
\caption{The representation of states in the five-qubit code, for a convenient definition of the Wigner function. a) The encoded computational state $\ket{0}_L$ is associated to the horizontal ray; b) the Wigner function for the state $\ket{+}_L =$ $= (\ket{0}_L + \ket{1}_L)/\sqrt{2}$ is formed by the lines associated to the encoded computational states, plus a pair of oscillating horizontal lines; all the encoded states will have a similar representation.}
\label{FIGwignercode}
\end{center}
\end{figure}

Now we can choose a quantum net such that the interference lies only on two horizontal lines $\vec p_{I}$ and $\vec p_{I} + \vec m$, for arbitrary $\vec p_I$. All we need to do is impose the conditions:
\begin{equation}
f'_{(\vec x, \vec m)} = \left\{ \begin{array}{ll}
(-1)^{\vec x \cdot \vec p_{I}} e^{i\frac{\pi}{2} \vec x \cdot \vec m}, & \vec x \cdot \vec m = 0 ~({\rm mod}~2) \\
-i (-1)^{\vec x \cdot \vec p_{I}} e^{i\frac{\pi}{2} \vec x \cdot \vec m}, & \vec x \cdot \vec m = 1 ~({\rm mod}~2)
\end{array} \right.
\end{equation}
(note that this can be done because $\vec m$ is fixed). Then the sum over all $n$-tuples $\vec x$ will be totally constructive at  $\vec p = \vec p_{I}$ and $\vec p = \vec p_{I} + \vec m$, and will vanish for all the other values of $\vec p$. The resulting Wigner function is given by (\ref{ECwignercode}), with:
\begin{eqnarray}
W_{I}'(\vec q, \vec p) & = & \frac{(-1)^{\vec q \cdot \vec m}}{d} \Big\{ \delta(\vec p - \vec p_{I}) ~[\mathcal{R}e (a b^*) + \mathcal{I}m (a b^*)] + \nonumber \\
&& + ~\delta(\vec p - \vec p_{I} - \vec m) ~[\mathcal{R}e (a b^*) - \mathcal{I}m (a b^*)] \Big\}
\end{eqnarray}
In Figure \ref{FIGwignercode} we show the representations of the graph-state $\ket{0}_L$, and the state $\ket{+}_L=(\ket{0}_L + \ket{1}_L)/\sqrt{2}$ that corresponds to the encoded version of the one-qubit eigenstate of $X$ with eigenvalue $+1$; all the encoded states have representations similar to these.

\smallskip

To conclude this section, let us analyse the representation of the state after a one-qubit Pauli error has occurred. 
By means of the discrete Wigner function we have defined, we can see that all one-qubit Pauli errors take the state to subspaces which are orthogonal to one another and to the subspace of encoded states. This is nothing but the condition for the error to be detectable and correctable by the code \cite{Shor95, Steane96}.
According to equation (\ref{ECwignerprod}), two states with Wigner functions $W_a' (\alpha)$ and $W_b' (\alpha)$ are orthogonal when $\sum_\alpha W'_a(\alpha) W'_b(\alpha) = 0$. 
As mentioned before, all the legal states have representations that correspond to constant non-negative values at the lines $\vec p = \vec 0$ and $\vec p = \vec m$, plus a couple of oscillating lines at $\vec p = \vec p_I$ and $\vec p = \vec p_I + \vec m$. The Wigner function vanishes for all the other values of $\vec p$. 

Because of the covariance condition (\ref{ECcovarianciatras}), the occurrence of a Pauli error translates the Wigner function in phase-space. For the translated state to be orthogonal to all encoded states, it suffices to see that the constant lines cannot be translated onto the lines $\vec p = \vec 0$ or $\vec p = \vec m$, and that the new oscillating lines cannot be placed at $\vec p = \vec p_I$ and $\vec p = \vec p_I + \vec m$. The coincidence of a translated constant (oscillating) line at the place of the original oscillating (constant) lines will not affect the orthogonality condition, as the sum over an oscillating line vanishes. This shows that the orthogonality between the encoded and the faulty states is granted if the translations associated to the errors sum to the $\vec p$ coordinate an $n$--tuple different from $\vec 0$ and $\vec m$.
Let us now consider how the possible one-qubit Pauli errors transform $\vec p$: 
\begin{enumerate}
\item $Z_{(j)}$ error: it is a vertical translation, that maps $p_j \to p_j +1$.
\item $X_{(j)}$ error: it is an oblique translation, equivalent to $S_j Z_{(j-1)} Z_{(j+1)}$; its vertical component changes $p_{j-1}$ and $p_{j+1}$. 
\item $Y_{(j)}$: it is also an oblique translation, equivalent to $S_j Z_{(j-1)} Z_{(j)} Z_{(j+1)}$, and so it affects $p_{j-1}$, $p_j$ and $p_{j+1}$.
\end{enumerate}
As we see, $Z$ errors transform only the component of $\vec p$ that corresponds to the faulty qubit, while $X$ errors change the components associated to the faulty qubit's neighbours, and $Y$ errors transform the components associated to the faulty qubit and its neighbours. None of these errors leaves $\vec p$ unchanged, and none changes the five components of $\vec p$. Then, the erroneous states are orthogonal to the subspace of encoded states. Besides, by comparison of the transformations of $\vec p$ it is easy to see that all one-qubit Pauli errors take the state to orthogonal subspaces, and so the different errors can be identified and corrected. Of course, this is no longer true when we consider Pauli errors that act on more than one qubit. For instance, an error of the form $Z_{(j-1)} Z_{(j+1)}$ cannot be distinguished from an error $X_{(j)}$, and so this code does not allow for the correction of two-qubit errors. 

\section{Conclusions} 
\label{sec:conclusion}

We studied the representation of interference in phase-space using the discrete Wigner functions defined in \cite{GibbonsHW04}. We showed that, for a coherent superposition of two stabilizer states, the interference fringes may spread over all phase space, even to the regions where the interfering states are localized. We analysed the genericity of this behaviour and concluded that for systems of two and three qubits there is no way to define the discrete Wigner function (by choosing a particular ``quantum net") that can avoid this overlap. We conjecture that the same result will hold for an arbitrary number of qubits. 
In the light of these results, we analysed the average behaviour of the interference term in the Wigner function of a coherent superposition of two fixed stabilizer states. We did this by picking random quantum nets, and studying the ratio between the maximal value of the interference term and the value of the Wigner function on the original lines; this ratio seems to decay exponentially with the number of qubits. Furthermore, the interference tends to spread taking absolute values uniformly distributed over all phase-space, as the number of qubits is increased. These features suggest that the use of Wigner functions in this class may not be appropriate to identify the presence of interference in phase-space. 

On the other hand, we showed that it is possible to define a Wigner function to represent a class of states in phase-space that includes the stabilizer quantum error-correcting codes. For this purpose it turned out convenient to define translation operators that are adapted to the code. We applied this idea to analyse the phase-space representation of the five-qubit error-correcting code. Not only we found a simple representation for all encoded states, but showed that single qubit errors also have a natural interpretation in this scenario, as they are phase space translations.

\end{document}